\def\rpb{{R.~P.~Brent}} %

\def\CACM{{\em Comm.\ ACM}}

\documentstyle[twocolumn]{article}

\makeatletter
\def\@normalsize{\@setsize\normalsize{12pt}\xpt\@xpt
\abovedisplayskip  8pt plus2pt minus5pt\belowdisplayskip \abovedisplayskip
\abovedisplayshortskip \z@ plus3pt\belowdisplayshortskip 6pt plus3pt
minus3pt\let\@listi\@listI}

\def\subsize{\@setsize\subsize{12pt}\xipt\@xipt}

\def\section{\@startsection {section}{1}{\z@}{18pt plus 2pt minus 2pt}
{9pt plus 2pt minus 2pt}{\large\bf}}

\def\subsection{\@startsection {subsection}{2}{\z@}{9pt plus 2pt minus 2pt}
{9pt plus 2pt minus 2pt}{\subsize\bf}}
\makeatother

\begin{document}

\date{}					%

\bibliographystyle{plain}

\title{\Large\bf Fast Normal Random Number Generators for Vector Processors%
\thanks{Copyright \copyright\ 1993, R.~P.~Brent.
Appeared as Technical Report TR-CS-93-04, CS Lab, ANU, March 1993.}}

\author{Richard P.\ Brent$^1$\\		%
	Computer Sciences Laboratory\\
	Australian National University\\
	Canberra, ACT 0200}

\maketitle
\thispagestyle{empty}

\subsection*{\centering Abstract}
{\em
We consider pseudo-random number generators suitable for vector
processors. In particular, we
describe vectorised implementations of the Box-Muller and Polar
methods, and show that they give good performance on the
Fujitsu VP2200. We also consider some other popular methods,
e.g.\ the Ratio method and the method of Von Neumann and Forsythe,
and show why they are unlikely to be competitive with the
Polar method on vector processors.
}

\section{Introduction}
\label{Sec:Intro}

Several recent papers~\cite{And90,Bre92,Pet92}
have considered the generation of uniformly distributed
pseudo-random numbers on vector and parallel computers.
In many applications, random numbers from specified non-uniform distributions
are required. These distributions may be
continuous (e.g.~normal, exponential) or
discrete (e.g.~Poisson).
A common requirement is for the normal distribution.

The most efficient methods for generating %
normally distributed random variables on sequential
machines~\cite{Ahr72,Bre74,Dev86,Gri85,Knu81,Lev92}
involve the use of different approximations
on different intervals, and/or the use of ``rejection'' methods,
so they often do not vectorise well.
Simple, ``old-fashioned'' methods
may be preferable. In Section~\ref{Sec:NRNG} we describe two such
methods, and in Sections~\ref{Sec:BOXM}--\ref{Sec:POLAR}
we consider their efficient implementation on vector processors,
and give the results of implementations on a Fujitsu VP2200/10.
In Sections~\ref{Sec:Ratio}--\ref{Sec:Forsythe}
we consider some other methods which are popular on serial machines,
and show that they are unlikely to be competitive on vector processors.

\section{Some Normal Generators}
\label{Sec:NRNG}

Assume that a good uniform random number generator which returns
uniformly distributed numbers in the interval $[0, 1)$ is available,
and that we wish to sample the normal distribution with
mean $\mu$ and variance~$\sigma^2$.
We can generate two independent,
normally distributed numbers $x$, $y$ by the following old algorithm
due to Box and Muller~\cite{Mul59} {\em (Algorithm~B1):}

\begin{itemize}
\item[1.] Generate independent uniform numbers $u$ and $v$.
\item[2.] Set $r \leftarrow \sigma\sqrt{-2\ln (1-u)}$.
\item[3.] Set $x \leftarrow r\sin(2\pi v) + \mu$
	  and $y \leftarrow r\cos(2\pi v) + \mu$.
\end{itemize}

\noindent Two minor points:
\begin{itemize}
\item[a.] We have written $(1-u)$ instead of $u$ at step 2 because of
our assumption that the uniform random number generator may
return exactly 0 but never exactly 1. If the uniform generator
never returns exactly 0, then $(1-u)$ can be replaced by~$u$.
Similar comments apply below.
\item[b.] The argument of sin and cos in step 3
is in the interval $[-\pi, \pi)$, but any interval of length $2\pi$
would be satisfactory.
\end{itemize}

\footnotetext[1]{E-mail address: {\tt rpb@cslab.anu.edu.au}\\
			\hspace*{\fill} rpb141tr typeset using \LaTeX}
\addtocounter{footnote}{1}	%
The proof that the algorithm is correct follows on
considering the distribution of $(x,y)$ transformed
to polar coordinates, and is similar to the proof of correctness
of the Polar method, given in~\cite{Knu81}.

Algorithm B1 is a reasonable choice if vectorised square root,
logarithm and trigonometric function routines are available.
Each normally distributed number requires
$1$ uniformly distributed number,
$0.5$ square roots,
$0.5$ logarithms, and
$1$ sin or cos evaluation.
Vectorised implementations of the Box-Muller method are discussed in
Section~\ref{Sec:BOXM}.

A variation of Algorithm B1 is the
{\em Polar} method of Box, Muller and Marsaglia~(1958)
({\em Algorithm~P} of Knuth~\cite{Knu81}):

\begin{itemize}
\item[1.] Generate independent uniform random numbers
$x$ and $y$ in the interval $[-1, 1)$.
\item[2.] Compute $s \leftarrow x^2 + y^2$.
\item[3.] If $s \ge 1$ then go to step 1
(i.e.~{\em reject} $x$ and $y$)
else go to step 4.
\item[4.] Set $r \leftarrow \sigma\sqrt{-2\ln(s)/s}$, %
and return $rx + \mu$ and $ry + \mu$.
\end{itemize}

It is easy to see that, at step~4,
$(x,y)$ is uniformly distributed in the unit circle,
so $s$ is uniformly distributed
in $[0, 1)$.
To avoid the remote possibility of division by zero at step~4,
we could replace $\ln(s)/s$ by $\ln(1-s)/(1-s)$.

A proof that the values returned by Algorithm~P are independent,
normally distributed random numbers (with mean $\mu$ and variance $\sigma^2$)
is given in Knuth~\cite{Knu81}.
On average, step 1 is executed $4/\pi$ times,
so each normally distributed number requires
$4/\pi \simeq 1.27$ uniform random numbers,
0.5 divisions, 0.5 square roots, and 0.5 logarithms.
Compared to Algorithm~B1, we have avoided the sin and cos computation
at the expense of more uniform random numbers, 0.5 divisions,
and the cost of implementing the acceptance/rejection process.
This can be done using a vector gather.
Vectorised implementations of the Polar method are discussed in
Section~\ref{Sec:POLAR}.

\section{Implementation of the Box-Muller Method}
\label{Sec:BOXM}

We have implemented the Box-Muller method (Algorithm~B1 above)
and several refinements (B2, B3) on a Fujitsu VP~2200/10 vector processor
at the Australian National University.
The implementations all return double-precision real
results, and in cases where approximations to sin, cos, sqrt
and/or ln have been made, the absolute error is considerably
less than $10^{-10}$. Thus, statistical tests using less than about
$10^{20}$ random numbers should not be able to detect any bias
due to the approximations.
The calling sequences allow for an array %
of random numbers to be returned. This permits vectorisation and amortises the
cost of a subroutine call over the cost of generating many
random numbers.

Our method~B2 is the same as B1, except that we replace calls to
the Fortran library sin and cos by an inline computation, using
a fast, but sufficiently accurate, approximation.
Let $y = (2\pi v - \pi)/16$, where $0 \le v < 1$, so $\vert y \vert \le \pi/16$.
We approximate $\sin y$ by
a polynomial of the form $s_1y + s_3y^3 + s_5y^5 + s_7y^7$,
and $\cos y$ by a polynomial of the form
$c_0 + c_2y^2 + c^4y^4 + c_6y^6$.
Then, using the identities
\[\sin 2y = 2\sin y \cos y, \;\;\; \cos 2y = 1 - 2\sin^2y \]
four times, we can compute $\sin 16y$ and $\cos 16y$ with a small number
of multiplications and additions. The computation is vectorizable.

Times, in machine cycles
per normally distributed number,
for methods B1, B2 (and other methods described below) are given
in Table~\ref{Tab:Box-times}.
In all cases the generalised Fibonacci random number generator
RANU4 (described in~\cite{Bre92}) was used to generate the required uniform
random numbers, and a large number of random numbers were
generated, so that vector lengths were long.
RANU4 generates a uniformly
distributed random number in 2.2 cycles on the VP~2200/10.
(The cycle time of the VP~2200/10 at ANU
is 3.2 nsec, and two multiplies and two adds can be performed per
clock cycle, so the peak speed is 1.25 Gflop.)

The Table gives the total times and also the estimated times for
the four main components:
\begin{itemize}
\item[1.] ln computation (actually 0.5 times the cost of one ln computation
since the times are per normal random number generated).
\item[2.] sqrt computation (actually 0.5 times).
\item[3.] sin or cos computation.
\item[4.] other, including uniform random number generation.
\end{itemize}

\begin{table}
\centerline{
\begin{tabular}{|c|c|c|c|c|c|c|} \hline
component	&B1	&B2	&B3	&P1	&P2	&R1\\ \hline
ln		&13.1	&13.1	&7.1	&13.1	&7.1	&0.3\\
sqrt		&8.8	&8.8	&1.0	&8.8	&1.0	&0.0\\
sin/cos		&13.8	&6.6	&6.6	&0.0	&0.0	&0.0\\
other		&5.9	&5.6	&11.6	&11.9	&13.8	&35.1\\ \hline
total		&41.6	&34.1	&26.3 	&33.8	&21.9	&35.4\\ \hline
\end{tabular}
}
\caption{Cycles per normal random number}
\label{Tab:Box-times}
\end{table}

The results for method~B1 show that the sin/cos and ln computations are
the most expensive (65\% of the total time).
Method~B2 is successful in reducing the sin/cos time from 33\% of
the total to 19\%.

In Method~B2, 64\% of the time is consumed by the computation
of $\sqrt{-\ln (1 - u)}$.
An obvious way to reduce this time is to use a fast approximation to
the function
\[ f(u) = \sqrt{-\ln (1 - u)} ,	\]
just as we used a fast approximation
to sin and cos to speed up method~B1. However, this is difficult to
accomplish with sufficient accuracy, because the function
$f(u)$ has singularities at both endpoints of the unit interval.
Method~B3 overcomes this difficulty in the following way.
\begin{itemize}
\item[1.] We approximate the function
\[ g(u) = u^{-1/2}f(u) = \sqrt{{-\ln (1 - u) \over u}} ,\]
rather than $f(u)$. Using the Taylor series for $\ln(1-u)$, we see
that $g(u) = 1 + u/4 + \cdots$ is well-behaved near $u = 0$.
\item[2.] The approximation to $g(u)$ is only used in the interval
$0 \le u \le \tau$, where $\tau < 1$ is suitably chosen.
For $\tau < u < 1$ we use the slow but accurate Fortran ln and sqrt routines.
\item[3.] We make a change of variable of the form
$ v = (\alpha u + \beta) / (\gamma u + \delta) $,
where $\alpha, \ldots, \delta$ are chosen
to map $[0, \tau]$ to $[-1, 1]$,
and the remaining degrees of freedom are used to move the singularities
of the function $h(v) = g(u)$ as far away as possible from the
region of interest (which is $-1 \le v \le 1$).
To be more precise, let $\rho$ be a positive parameter.
Then we can choose
\[ \tau = 1 - \left({\rho \over \rho + 2}\right)^2 ,\]
\[ v = (\rho + 1)\left({(\rho + 2)u - 2 \over
		2(\rho + 1) - (\rho + 2)u}\right) ,\]
and the singularities of $h(v)$ are at $\pm(\rho + 1)$.
\end{itemize}

For simplicity, we choose $\rho = 1$, which experiment shows is close to
optimal on the VP~2200/10. Then $\tau = 8/9$,
$ v = (6u - 4) / (4 - 3u) $,
and $h(v)$ has singularities at $v = \pm 2$, corresponding to
the singularities of $g(u)$ at $u = 1$ and $u = \infty$.
A polynomial
of the form $h_0 + h_1v + \cdots + h_{15}v^{15}$
can be used to approximate $h(v)$ with absolute error less than
$2 \times 10^{-11}$ on $[-1, 1]$.
About 30 terms would be needed if we attempted to approximate
$g(u)$ to the same accuracy by a polynomial on $[0, \tau]$.
We use polynomial approximations which are close
to minimax approximations. These
may easily be obtained by truncating Chebyshev series,
as described in~\cite{Cle61}.

It appears that this approach requires the computation of a
square root, since we really want $f(u) = u^{1/2}g(u)$,
not $g(u)$. However, a trick allows this square root
computation to be avoided, at the expense of an additional
uniform random number generation (which is cheap) and a
few arithmetic operations.  Recall that $u$ is a uniformly
distributed random variable on $[0, 1)$.  We generate {\em two}
independent uniform variables, say $u_1$ and $u_2$,
and let $u \leftarrow \max(u_1, u_2)^2$.
It is easy to see that $u$ is in fact uniformly distributed
on $[0, 1)$. However,
$ u^{1/2} = \max(u_1, u_2) $
can be computed without calling the library sqrt routine.
To summarise, a non-vectorised version of method~B3 is:

\begin{itemize}
\item[1.] Generate uniform random numbers $u_1$, $u_2$ and $u_3$ on $[0, 1)$.
\item[2.] Set $m \leftarrow \max(u_1, u_2)$ and $u \leftarrow m^2$.
\item[3.] If $u > 8/9$ then
\item[]3.1. set $r \leftarrow \sigma\sqrt{-\ln(1-u)}$ using Fortran
		library routines, else
\item[]3.2. set $v \leftarrow (6u-4)/(4 - 3u)$,
	    evaluate $h(v)$ as described above,
	    and set $r \leftarrow \sigma mh(v)$.
\item[4.] Evaluate $s \leftarrow \sin(2\pi u_3 - \pi)$
	and $c \leftarrow \cos(2\pi u_3 - \pi)$ as described above.
\item[5.] Return $\mu + cr\sqrt{2}$ and $\mu + sr\sqrt{2}$,
         which are independent, normal random
	numbers with mean $\mu$ and standard deviation $\sigma$.

\end{itemize}

Vectorization of method~B3 is straightforward, and can take
advantage of the ``list vector'' technique on the VP2200.
The idea is to gather those $u > 8/9$ into a contiguous array,
call the vectorised library routines to compute an array of
$\sqrt{-\ln(1-u)}$ values, and scatter these back.
The gather and scatter operations do introduce some
overhead, as can be seen from the row labelled ``other'' in the Table.
Nevertheless, on the VP2200,
method~B3 is about 23\% faster than method~B2, and
about 37\% faster than the straightforward method~B1.
These ratios could be different on machines with more (or less)
efficient implementations of scatter and gather.

Petersen~\cite{Pet92} gives times for normal and uniform
random number generators on a NEC SX-3.
His implementation {\em normalen} of the Box-Muller method takes 55.5 nsec per
normally distributed number, i.e.~it is 2.4 times faster than
our method~B1, and 1.51 times faster than our method~B3.
The model of SX-3 used by Petersen has an effective peak speed of
2.75 Gflop,	% 5.5 Gflop with 2 processors, but he only uses 1 processor
which is 2.2 times the peak speed of the VP~2200/10.
Considering the relative speeds
of the two machines and the fact that the SX-3
has a hardware square root function, our results are quite
encouraging.

\section{Implementation of the Polar Method}
\label{Sec:POLAR}

The times given in Table~\ref{Tab:Box-times} for methods B1--B3
can be used to predict
the best possible performance of the Polar method
(Section~\ref{Sec:NRNG}). The Polar method avoids the computation of
sin and cos, so could gain up to 6.6 cycles per normal random number
over method~B3. However, we would expect the gain to be less
than this because of the overhead of a vector gather
caused by use of a rejection method.
A straightforward vectorised implementation of the Polar method,
called method~P1, was written to test this prediction.
The results are shown in Table~\ref{Tab:Box-times}.
13.8 cycles are saved by avoiding
the sin and cos function evaluations, but the overhead increases
by 6.0 cycles, giving an overall saving of 7.8 cycles or 19\%.
Thus, method~P1 is about the same speed as method~B2, but not as
fast as method~B3.

Encouraged by our success in avoiding most ln and sqrt computations
in the Box-Muller method (see method~B3), we considered if
a similar idea would work for the Polar method.
In fact, it does. Step 4 of the Polar method (Section~\ref{Sec:NRNG})
involves the computation of $\sqrt{-2\ln(s)/s}$, where
$0 < s < 1$. The function has a singularity at $s = 0$, but we can
approximate it quite well on an interval such as $[1/9, 1]$,
using a method similar to that used to approximate the
function $g(u)$ of Section~\ref{Sec:BOXM}.

Inspection of the proof in Knuth~\cite{Knu81} shows that step 4 of
the Polar method can be replaced by

\begin{itemize}
\item[4a.] Set $r \leftarrow \sigma\sqrt{-2\ln(u)/s}$,\\ % (s) or (1-s) OK
and return $rx+\mu$ and $ry+\mu$
\end{itemize}
where $u$ is any uniformly distributed variable on $(0, 1]$,
provided $u$ is independent of $\arctan (y/x)$.
In particular, we can take $u = 1 - s$. Thus, omitting the constant
factor $\sigma\sqrt{2}$, we need to evaluate $\sqrt{-\ln(1-s)/s}$,
but this is just $g(s)$, and we can use exactly the same approximation
as in Section~\ref{Sec:BOXM}. This gives us method~P2.
To summarise, a non-vectorised version of method~P2 is:

\begin{itemize}
\item[1.] Generate independent uniform random numbers
$x$ and $y$ in the interval $[-1, 1)$.
\item[2.] Compute $s \leftarrow x^2 + y^2$.
\item[3.] If $s \ge 1$ then go to step 1
(i.e.~{\em reject} $x$ and $y$)
else go to step 4.
\item[4.] If $s > 8/9$ then
\item[]4.1. set $r \leftarrow \sigma\sqrt{-\ln(1-s)/s}$ using Fortran
		library routines, else
\item[]4.2. set $v \leftarrow (6s-4)/(4-3s)$,
	    evaluate $h(v)$ as described in Section~\ref{Sec:BOXM},
	    and set $r \leftarrow \sigma h(v)$.
\item[5.] Return $xr\sqrt{2} + \mu$ and $yr\sqrt{2} + \mu$,
         which are independent, normal random
	numbers with mean $\mu$ and standard deviation $\sigma$.
\end{itemize}

To vectorise steps 1-3, we simply generate vectors of $x_j$ and $y_j$
values, compute $s_j = x_j^2 + y_j^2$, and compress by omitting
any triple $(x_j, y_j, s_j)$ for which $s_j \ge 1$.
This means that we can not predict in advance how many normal
random numbers will be generated, but this problem is easily
handled by introducing a level of buffering.
The vectorised version of method~P2 is called RANN3B,
and the user-friendly routine which performs the buffering
and calls RANN3B is called RANN3.

The second-last column of Table~\ref{Tab:Box-times} gives results for
method~P2 (actually for RANN3, since the buffering overhead is
included). There is a saving of 11.9 cycles or 35\% compared
to method~P1, and the method is 17\% faster than the fastest
version of the Box-Muller method (method~B3).  The cost of
logarithm and square root computations is only 37\% of the total,
the remainder being the cost of generating uniform random
numbers (about 13\%) and the cost of the rejection step
and other overheads (about 50\%). On the VP2200/10 we can
generate more than 14 million normally distributed random numbers
per second (one per 70 nsec).

\section{The Ratio Method}
\label{Sec:Ratio}

The Polar method is one of the simplest of a class of rejection
methods for generating random samples from the normal (and other)
distributions.
Other examples are given in~\cite{Ahr72,Bre74,Dev86,Knu81}.
It is possible to implement some
of these methods in a manner similar to our implementation of
method~P2.
For example, a popular method is the Ratio Method of Kinderman
and Monahan~\cite{Kin77} (also described in~\cite{Knu81},
and improved in~\cite{Lev92}).
In its simplest form, the Ratio Method is given by
{\em Algorithm~R:}

\begin{itemize}
\item[1.] Generate independent uniform random numbers $u$ and $v$
in $[0, 1)$.
\item[2.] Set $x \leftarrow \sqrt{8/e}(v-{1 \over 2})/(1-u)$.
\item[3.] If $-x^2\ln(1-u) > 4$ % NB ln(1-u) <= 0
then go to step 1 (i.e.\ reject~$x$)
	else go to step 4.
\item[4.] Return $\sigma x + \mu$.
\end{itemize}

Algorithm~R returns a normally distributed random number using
(on average) $8/\sqrt{\pi e} \simeq 2.74$ uniform random numbers
and 1.37 logarithm evaluations.
The proof of correctness,
and various refinements which reduce the number of logarithm
evaluations, are given in~\cite{Kin77,Knu81,Lev92}.
The idea of the proof is that $x$ is normally distributed
if the point $(u,v)$ lies inside a certain closed curve
$C$ which in turn is inside the rectangle
$[0,1] \times [-\sqrt{2/e},+\sqrt{2/e}]$.
Step~3 rejects $(u,v)$ if it is outside $C$.

The function $\ln(1-u)$ occurring at step~3 has a singularity
at $u = 1$, but it can be evaluated using a polynomial or rational
approximation on some interval $[0, \tau]$, where $\tau < 1$,
in much the same way as the function $g(u)$ of
Section~\ref{Sec:BOXM}.

The refinements added by Kinderman and Monahan~\cite{Kin77}
and Leva~\cite{Lev92} avoid most of the logarithm evaluations.
The following step is added:

\begin{itemize}
\item[2.5.] If $P_1(u,v)$ then go to step 4\\
else if $P_2(u,v)$ then go to step 1\\
else go to step 3.
\end{itemize}

Here $P_1(u,v)$ and $P_2(u,v)$ are easily-computed conditions.
Geometrically, $P_1$ corresponds to a region $R_1$ which
lies inside $C$,
and $P_2$ corresponds to a region $R_2$ which encloses $C$,
but $R_1$ and $R_2$ have almost the same area.
Step~3 is only executed if $(u,v)$ lies in
the borderline region $R_2\backslash R_1$.

Step~2.5 can be vectorised, but at the expense of several vector
scatter/gather operations. Thus, the saving in logarithm evaluations
is partly cancelled out by an increase in overheads.
The last column (R1) of Table~\ref{Tab:Box-times} gives the times
for our implementation on the VP2200. As expected, the time for the logarithm
computation is now negligible, and the overheads dominate.
In percentage terms the times are:
\begin{itemize}
\item[1\%] logarithm computation (using the library routine),
\item[17\%] uniform random number computation,
\item[23\%] scatter and gather to handle borderline region,
\item[59\%] step~2.5 and other overheads.
\end{itemize}
Although disappointing, the result for the Ratio method is not
surprising, because the computations and overheads are similar
to those for method~P2 (though with less logarithm computations),
but only half as many normal random numbers are produced.
Thus, we would expect the Ratio method to be slightly better than
half as fast as method~P2, and this is what Table~\ref{Tab:Box-times} shows.

\section{GRAND}
\label{Sec:Forsythe}

On serial machines GRAND~\cite{Bre74} is competitive
with the Ratio method. In fact, GRAND is the fastest of the
methods compared by Leva~\cite{Lev92}. GRAND is based on an
idea of Von Neumann and Forsythe for generating
samples from a distribution with density function
$c\exp(-h(x))$, where $0 \le h(x) \le 1$:
\begin{enumerate}
\item Generate a uniform random number $x$,
and set $u_0 \leftarrow h(x)$.
\item Generate independent uniform random
numbers $u_1, u_2, \ldots$\\
until the first $k > 0$ such that $u_{k-1} < u_k$.
\item If $k$ is odd then return $x$,\\
 else reject $x$ and go to step~1.
\end{enumerate}
A proof of correctness is given in Knuth~\cite{Knu81}.
\medskip

It is hard to see how to implement GRAND efficiently
on a vector processor. There are two problems~--

\begin{enumerate}
\item $k$ is not bounded,
even though its expected value is small.
Thus, a sequence of gather operations seems to be required.
The result would be similar to Petersen's implementation~\cite{Pet92}
of a generator for the Poisson distribution (much slower
than his implementation for the normal distribution).
\item Because of the restriction $0 \le h(x) \le 1$, the area under
the normal curve $\exp(-x^2/2)/\sqrt{2\pi}$
has to be split into different regions
from which samples are drawn with probabilities proportional to their
areas.  This complicates the implementation of the rejection step.
\end{enumerate}

For these reasons we would expect a vectorised implementation of GRAND
to be even slower than our implementation of the Ratio method.
Similar comments apply to other rejection methods which use
an iterative rejection process and/or several different regions.

\section{Conclusion}
\label{Sec:Conc}

We have shown that both the Box-Muller and Polar methods vectorise
well, and that it is possible to avoid and/or speed up the evaluation
of the functions (sin, cos, ln, sqrt) which appear necessary.
On the VP2200/10 our best implementation of the Polar method
takes 21.9 machine cycles per normal random number,
slightly faster than our best implementation of the Box-Muller
method (26.3 cycles).

We also considered the vectorisation of some other popular methods
for generating normally distributed random numbers,
such as the Ratio method and the method of Von Neumann and Forsythe,
and showed why such methods
are unlikely to be faster than the Polar method on a vector
processor.

\subsection*{Acknowledgements}				%

Thanks to Dr~W.~Petersen for his comments and helpful information on
implementations of random number generators on
Cray and NEC computers~\cite{Pet88,Pet92,Pet93}.
The ANU Supercomputer Facility provided time on the VP~2200/10 for
the development and testing of our implementation.
This work was supported in part by a Fujitsu-ANU research agreement.


\begin{thebibliography}{99}

\bibitem{Ahr72}
J.~H.~Ahrens and U.~Dieter,
``Computer methods for sampling from the exponential and normal
distributions'',
{\em Communications of the ACM} 15 (1972), 873-882.

\bibitem{And90}
S.~L.~Anderson,
``Random number generators on vector supercomputers and other advanced
architectures'',
{\em SIAM Review} 32 (1990), 221-251.

\bibitem{Bre74} \rpb, Algorithm 488: A Gaussian pseudo-random number
generator (G5),
\CACM\ 17 (1974), 704-706. %

\bibitem{Bre92}
R.~P.~Brent,
Uniform random number generators for supercomputers,
{\em Proc.\ Fifth Australian Supercomputer Conference},
Melbourne, December 1992, 95-104.

\bibitem{Cle61}	%
C.~W.~Clenshaw,
L.~Fox, E.~T.~Goodwin, D.~W.\ Martin,
J.~G.\ L.\ Michel, G.~F.~Miller, F.~W.\ J.\ Olver and J.~H.~Wilkinson,
{\em Modern Computing Methods}, 2nd edition, HMSO, London, 1961, Ch.~8.

\bibitem{Dev86}
    L.~Devroye,
    {\em Non-Uniform Random Variate Generation.}
    Springer-Verlag, New York, 1986.

\bibitem{Gri85}
P.~Griffiths and I.~D.~Hill (editors),
{\em Applied Statistics Algorithms},
Ellis Horwood, Chichester, 1985.

\bibitem{Jam90}
F.~James,
A review of pseudorandom number generators.
{\em Computer Physics Communications}, 60:329--344, 1990.

\bibitem{Kin77}
A.~J.~Kinderman and J.~F.~Monahan,
``Computer generation of random variables using the ratio of
uniform deviates'',
{\em ACM Transactions on Mathematical Software} 3 (1977), 257-260.

\bibitem{Knu81}
    D.~E.~Knuth,
    {\em The Art of Computer Programming,
    Volume 2: Seminumerical Algorithms} (second edition).
    Addison-Wesley, Menlo Park, 1981, Sec.\ 3.4.1.

\bibitem{Lev92}
J.~L.~Leva, ``A fast normal random number generator'',
{\em ACM Transactions on Mathematical Software} 18 (1992), 449-453.

\bibitem{Mar90a}
    G.~Marsaglia, B.~Narasimhan and A.~Zarif,
    A random number generator for PC's.		% "PC's", not "PCs" !
    {\em Computer Physics Communications}, 60:345--349, 1990.

\bibitem{Mar91}
    G.~Marsaglia and A.~Zaman,
    A new class of random number generators.
    {\em The Annals of Applied Probability}, 1:462--480, 1991.

\bibitem{Mul59}
    M.~E.~Muller,
    A comparison of methods for generating normal variates on
    digital computers.
    {\em J.~ACM} 6:376--383, 1959.

\bibitem{Pet88}	%
    W.~P.~Petersen,
    Some vectorized random number generators for uniform,
    normal, and Poisson distributions for CRAY X-MP,
    {\em J.~Supercomputing}, 1:327--335, 1988.

\bibitem{Pet92}
    W.~P.~Petersen,
    {\em Lagged Fibonacci Series Random Number Generators for the
    NEC SX-3},
    IPS Research Report No.~92-08,
    IPS, ETH-Zentrum, Zurich, April 1992.

\bibitem{Pet93}
    W.~P.~Petersen,
    {\em Random Number Generators on Vector Architectures},
    preprint, 1993.

  \end{thebibliography}
\end{document}